\documentclass{article}
\usepackage{spconf}

\usepackage[utf8]{inputenc}
\usepackage[T1]{fontenc}
\usepackage{amsmath,amssymb}
\usepackage{graphicx}
\usepackage{booktabs}
\usepackage{multirow}
\usepackage{microtype}
\usepackage{xcolor}
\usepackage{url}
\usepackage[hidelinks]{hyperref}
\definecolor{pendingred}{RGB}{165,25,25}
\definecolor{resultgreen}{RGB}{20,110,65}

\newcommand{\best}[1]{\textcolor{resultgreen}{\textbf{#1}}}
\newcommand{\model}{CosyVoice3}
\newcommand{\corpus}{Balalaika-Longform}

\urldef{\datasetlink}\url{https://huggingface.co/datasets/lab260/Balalaika-longform}
\urldef{\codelink}\url{https://github.com/lab260ru/balalaika-longform}
\newcommand{\codestatement}{Training, inference and evaluation code, the evaluation
  pairs with their 21 reference prompts, and per-attempt transcripts, alignment
  counts, stop reasons and window-level scores are released.\footnote{\codelink}}
\newif\ifworkingdraft
\workingdraftfalse 

\title{Balalaika-Longform: A Russian Speech Corpus\\
for Continuous Long-Form Text-to-Speech}

\name{Nikita Vasiliev$^{1,2}$, Kirill Borodin$^{1,2}$, Vasilii Kudryavtsev$^{1,2}$, Maxim Maslov$^{2}$, Grach Mkrtchian$^{2,3}$}
\address{$^{1}$BitmanagerAI, Dubai, UAE\quad
  $^{2}$lab260, Yerevan, Armenia\quad
  $^{3}$MTUCI, Moscow, Russia\\
  \texttt{kborodin.research@gmail.com}}

\usepackage{eso-pic}
\AddToShipoutPictureFG*{%
  \put(0,34){\makebox[\paperwidth]{\parbox{6.4in}{\centering\footnotesize
    This work has been submitted to the IEEE ICASSP for possible publication.
    Copyright may be transferred without notice, after which this version
    may no longer be accessible.}}}}

\begin{document}
\ninept
\maketitle

\begin{abstract}
Long-form text-to-speech must retain requested words over extended generations,
yet sentence-level training and evaluation can hide omissions and early stops.
We introduce \corpus{}, an open Russian corpus of 189 hours in
continuous units of 30 seconds to 15 minutes. Long units and matched short windows support fine-tuning comparisons on the same source
recordings, and the accompanying evaluation retains every synthesis attempt. We fine-tune
CosyVoice3, Qwen3-TTS, VoxCPM2 and F5-TTS on either view and test
continuous synthesis on 50 text--voice pairs with voices unseen in
fine-tuning, at about 75, 300 and 1,200 words. At 1,200 words, CosyVoice3 WER
falls from 99.9\% after short-window fine-tuning to 47.3\% with long targets, and to 16.6\% with punctuated, format-matched transcripts; paired intervals
favor long targets for Qwen3-TTS and VoxCPM2, and by a small margin for F5-TTS,
whose absolute WER stays above 90\%. The results isolate the effect of training sequence length under a fixed
continuous-generation protocol.

\end{abstract}

\begin{keywords}
text-to-speech, Russian speech, speech corpus, long-form synthesis
\end{keywords}

\section{Introduction}
\label{sec:intro}

Audiobooks, lectures and podcasts require speech synthesis to preserve content
and voice beyond a sentence. Short outputs can conceal omissions, repetitions
and early stops, and listeners judge sentences and paragraphs differently
\cite{clark2019evaluating}. Existing pipelines obtain long audio by splitting
text into short generations~\cite{qwen2025cosyvoicecode,chen2026f5code}, but
splitting discards cross-sentence prosodic and discourse context and cannot test
whether a model can sustain a reading by itself. Studying that capability requires
data and evaluation that retain long context.  Chunked synthesis is a different
operating regime and outside our scope: we study how models behave when a long
text is given in a single pass.

We introduce \corpus{} for this purpose in Russian. The Balalaika
pipeline~\cite{borodin2026balalaika} supplies context-preserving segmentation,
transcripts, provenance and quality screening; the corpus retains continuous
speech units up to 15 minutes. Paired short-window and long-sequence training
views allow a source-controlled comparison of sequence length, with channel-disjoint
train, development and test splits.

We evaluate these views on CosyVoice3, Qwen3-TTS, VoxCPM2 and F5-TTS. A controlled
two-voice CosyVoice3 experiment tests duration under matched training exposure
(Table~\ref{tab:bybucket}); a published-text comparison of 50 fixed text--voice
pairs from 22 works (Table~\ref{tab:crossmodel}) retains every failure in
word-error and correct-word-recall estimates, with paired Short/Long intervals
for all four backbones.

Our contributions are the long-form corpus, paired training views and an
all-attempt evaluation of corpus use across generation families. The experiments
separate this data question from the choice to generate continuously or in chunks;
transcript-source, formatting and corpus-composition confounds are stated explicitly.

\section{Related Work}
\label{sec:related}

\noindent\textbf{Speech corpora for synthesis.}
LibriTTS-R improves the acoustic quality of an existing English corpus through speech
restoration~\cite{koizumi2023librittsr}, HiFiTTS-2 scales audiobook-derived data
with quality filtering and detailed metadata~\cite{langman2025hifitts2}, and
Libriheavy keeps punctuation, casing and preceding context for 50,000 hours of
audiobook speech~\cite{kang2024libriheavy}.
For Russian, RUSLAN provides carefully recorded single-speaker
speech~\cite{gabdrakhmanov2019ruslan}, and Balalaika provides a pipeline for
context-preserving segmentation, multi-ASR transcription, and speaker-purity
screening~\cite{borodin2026balalaika}.  We build on the latter to produce a corpus
organized around long continuous speech units with paired short and long training views.

\noindent\textbf{Approaches to long-form generation.}
Recent work addresses long speech through inference and model design.
MagpieTTS-LF carries state and text history across sentence chunks without
retraining~\cite{ghosh2026magpiettslf}.  VoiceStar combines positional modeling and a
training scheme for duration control and extrapolation~\cite{peng2026voicestar}.
SpeechSSM samples multi-minute speech and introduces the LibriSpeech-Long
benchmark for English~\cite{park2025speechssm}.
F5-TTS offers a non-autoregressive flow-matching
alternative~\cite{chen2024f5tts}; its standard long-text inference splits the text
and joins separately synthesized chunks~\cite{chen2026f5code}.
Our study instead examines the contribution of training data: whether preserving
long targets helps existing models under a fixed uninterrupted-inference protocol,
without assuming that long-form training is necessary for every synthesis method.

\noindent\textbf{Backbones and evaluation.}
CosyVoice~\cite{du2024cosyvoice2,du2025cosyvoice3},
Qwen3-TTS~\cite{hu2026qwen3tts}, and VoxCPM2~\cite{zhou2026voxcpm2}
provide open autoregressive models with different speech representations; we use
their public fine-tuning paths to assess the same corpus across backbones.
LFSBench evaluates long-form speech across scenarios and dimensions including
content, acoustics, and expressiveness~\cite{pan2026lfsbench}.  Russian LibriSpeech~\cite{openslr96ruls} releases utterance-level audiobook
segments, and the multilingual YODAS crawl~\cite{li2023yodas} includes unsegmented
Russian YouTube audio but is not curated for single-speaker synthesis.  To our
knowledge, \corpus{} is the first openly released Russian corpus curated for
long-form TTS, and its all-attempt evaluation the first public long-form TTS
evaluation for Russian.
Our evaluation addresses a narrower training-data question: it controls input
duration, retains failed attempts in the reported metrics, and evaluates whether a
single generation reads the requested text.

\section{Corpus and Evaluation Design}
\label{sec:method}

\subsection{Corpus, data splits, and training arms}

\corpus{} v3.1 is released openly as WebDataset shards.\footnote{\datasetlink} The open-source Balalaika pipeline~\cite{borodin2026balalaika} provided context-aware
segmentation, multi-ASR ROVER transcription, and quality and speaker-purity screening.
After loudness normalization, MossFormer2 speech
enhancement~\cite{zhao2024mossformer2} attenuated applause and background noise; the release and all training use this enhanced audio, and originals remain
locatable through the stored source URL and time span.

The corpus contains 6,380 single-speaker Russian segments (189.13 h; 30--900 s)
from 313 YouTube videos on 41 channels (seven hold 84\% of the hours), mainly
audiobook readings, lectures and talks, each carrying an uploader-declared Creative Commons Attribution flag,
which is license evidence rather than a rights audit.  Audio is
16-bit FLAC at the source rate (44.1 or 48 kHz, mostly stereo).  Each segment
carries a punctuated, capitalized GigaAM-v3-e2e-CTC~\cite{kutsakov2025gigaam}
transcript; where available, the recognizer-output voting
(ROVER)~\cite{fiscus1997rover} consensus text over ASR hypotheses; one of 462
video-local diarization labels; and its source time span, URL and license evidence.
Transcripts and speaker labels are automatic and uncorrected.  Unit length is skewed: 3,538 units last 30--60 s, 1,733 last 1--2 min, 699 last
2--5 min, 276 last 5--10 min and 134 last 10--15 min, so the median is 46 s and
the mean 107 s, yet units of at least 2 min hold 54\% of the hours and the 410 units of
at least 5 min hold 34\%.  We partition by YouTube channel before windowing: all
segments of a channel belong to one split (Table~\ref{tab:splits}); held-out
units are longer on average (dev 255 s, test 169 s, train 102 s).  All supervised
fine-tuning (SFT) uses train only; dev supplies validation and the two-voice set's
texts, and dev and test channels, never used for gradient updates, supply
reference voices.  Channel identity is a conservative document/speaker proxy,
supplemented by exact-text and 8-gram leakage checks and a speaker-embedding
gate across splits.

\begin{table}[!ht]
    \caption{Channel-disjoint splits before training-text exclusions. The remaining
  24 segments (0.88 h) come from seven held-out videos, one otherwise unrepresented,
  that fail the ASR-consistency gate; they are released separately and never used.}
  \label{tab:splits}
  \centering
  \small
  \setlength{\tabcolsep}{4pt}
  \begin{tabular}{lrrrr}
    \toprule
    Split & Segments & Hours & Videos & Channels\\
    \midrule
    Train & 6,043 & 170.68 & 265 & 27\\
    Dev   & 120   & 8.50   & 17 & 7\\
    Test  & 193   & 9.07   & 30 & 7\\
    \midrule
    Corpus & 6,380 & 189.13 & 313 & 41\\
    \bottomrule
  \end{tabular}
\end{table}

Training starts from the 6,043 released long units or 28,532 short windows that
tile the same recordings (contiguous, at most 30 s, mean 21.5 s; 170.68 h,
1,257,944 words before per-view text exclusions); window boundaries and aligned
transcripts are released as manifests with the code rather than as audio.  For \model{},
each long speech-token sequence concatenates its windows.  Fine-tuning on the
windows is Short-SFT and on the long units Long-SFT, both with ROVER text;
Long-SFT-Punct keeps the Long-SFT audio tokens but substitutes
punctuated, capitalized GigaAM-v3-e2e-CTC transcripts.  This also changes words: 95.0\% of the 6,041 training transcript pairs differ
after evaluation normalization, with 5.4\% pooled word-edit disagreement, so this
is not a punctuation-only ablation.

\subsection{Fine-tuning to assess corpus utility}
\label{sec:sft}

The main comparison changes training-sequence length while controlling exposure to
the corpus.  All \model{} arms start from the same checkpoint and fine-tune only the
language model; the speech tokenizer, acoustic generator, and speaker encoder stay
fixed.  They use 3,000 updates with the same learning rate and loss normalization.
Relative to Long-SFT, observed target-token exposure differs by 1.40\% for Short-SFT
and by 1.27\% for Long-SFT-Punct.

We fine-tune Qwen3-TTS and VoxCPM2 with their public training
code~\cite{qwen2026qwen3ttscode,openbmb2026voxcpmcode}, using punctuated text in both
arms (their Short-SFT and Long-SFT-Punct).  Within-backbone exposure differs by
0.17\% for Qwen3-TTS and 0.30\% for VoxCPM2.  VoxCPM2's training filter drops the
144 long units whose packed token sequences exceed its 8,192-position budget
(517--899 s; 18.2\% of the audio; retained units reach 661 s), while its short arm
keeps windows from these recordings, so that comparison does not fully isolate
duration from corpus composition.  The \model{} and Qwen3-TTS arms
share recording sets.

For F5-TTS, both arms fine-tune the full community Russian F5-TTS v1 Base
checkpoint~\cite{misha2025f5ru} with the flow-matching objective and punctuated
transcripts.  They traverse the same 6,038 recordings in
the same order for three passes (18,114 updates), using intact recordings or short
windows.  Long/Short consume 510.92/510.26 hours of training audio, within 0.24\% of
the Qwen3-TTS arms' audio budgets; final exponential-moving-average (EMA) weights are evaluated without
score-based checkpoint selection.  Both arms exclude records lacking usable windows;
residual normalized transcript disagreement is 0.04\%.  \codestatement{}

\begin{table*}[t]
    \caption{\model{} results on the earlier two-voice set (six texts, five
  duration buckets, two held-out voices; 60 attempts per arm) by input duration;
  all synthesis attempts are retained. Best value per row in bold.}
      \label{tab:bybucket}
  \centering
  \small
  \renewcommand{\arraystretch}{0.86}
  \setlength{\tabcolsep}{3.0pt}
  \begin{tabular}{llrrrr@{\hspace{14pt}}rrrr}
    \toprule
    & & \multicolumn{4}{c}{Complete $\uparrow$ (\%)}
      & \multicolumn{4}{c}{All-attempt WER $\downarrow$ (\%)}\\
    \cmidrule(lr){3-6}\cmidrule(lr){7-10}
    Bucket & Target duration
    & Base & Short-SFT & Long-SFT & Long-SFT-Punct
    & Base & Short-SFT & Long-SFT & Long-SFT-Punct\\
    \midrule
    B0 & 20--40 s & 83.3 & 91.7 & \best{100.0} & \best{100.0} & 7.18 & 7.07 & 3.45 & \best{2.93}\\
    B1 & 60--90 s & 58.3 & 41.7 & \best{83.3}  & 75.0 & 14.56 & 17.04 & 11.70 & \best{9.61}\\
    B2 & 2--3 min & 0.0 & 0.0 & \best{100.0} & \best{100.0} & 90.86 & 82.64 & \best{2.74} & 3.42\\
    B3 & 4--6 min & 0.0 & 0.0 & 41.7 & \best{75.0} & 98.07 & 97.77 & 47.22 & \best{36.66}\\
    B4 & 8--12 min & 0.0 & 0.0 & 33.3 & \best{66.7} & 99.82 & 99.89 & 37.22 & \best{19.15}\\
    \midrule
    All & 20 s--12 min & 28.3 & 26.7 & 71.7 & \best{83.3} & 62.10 & 60.88 & 20.47 & \best{14.36}\\
    \bottomrule
  \end{tabular}
\end{table*}

\begin{figure*}[t]
\centering
\includegraphics[width=\textwidth]{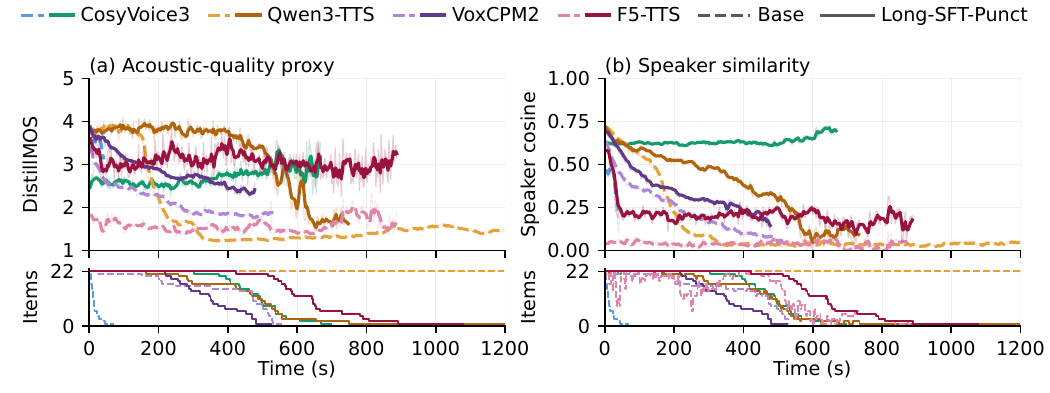}
\caption{Windowed diagnostics on the same 22 longest inputs per condition; Base
and Long-SFT-Punct are shown, and all 13 conditions accompany the code release.
Both metrics use 5-s windows with a 2.5-s hop; speaker windows require at least 1 s
of voiced audio. Faint curves are item-weighted means, bold curves exponential
smoothing ($\alpha=0.3$); lower strips count contributing items (at least two per
score).}
\label{fig:qualitytime}
\end{figure*}

\begin{table*}[t]
\caption{Published-text evaluation of all 13 continuous conditions by input length (14, 14 and 22 attempts per condition; texts from 22 works, 21 voices unseen in fine-tuning). All-attempt WER and correct-word recall include failed outputs; brackets are work-cluster bootstrap 95\% intervals; text differences use unrounded means; best point estimate per backbone and column in bold (not a significance statement). Qwen3-TTS, VoxCPM2 and F5-TTS also use punctuated text for Short-SFT.}
\label{tab:crossmodel}
\centering
\small
\renewcommand{\arraystretch}{0.88}
\setlength{\tabcolsep}{5pt}
\begin{tabular}{llrrrr}
\toprule
& & \multicolumn{3}{c}{All-attempt WER $\downarrow$ (\%) [95\% CI] by input length} & Recall $\uparrow$ (\%)\\
\cmidrule(lr){3-5}\cmidrule(lr){6-6}
Backbone & Arm & $\sim$75 words & $\sim$300 words & $\sim$1,200 words & $\sim$1,200 words\\
\midrule
CosyVoice3 & Base & 7.3 [4.1, 11.2] & 84.8 [77.9, 90.9] & 99.8 [99.7, 99.9] & 0.2\\
 & Short-SFT & 7.7 [4.9, 10.8] & 82.2 [72.5, 90.3] & 99.9 [99.9, 100.0] & 0.1\\
 & Long-SFT & 5.2 [2.8, 8.0] & 13.8 [5.7, 24.1] & 47.3 [37.2, 57.7] & 69.5\\
 & Long-SFT-Punct & \best{3.6} [2.2, 5.0] & \best{5.6} [2.7, 9.8] & \best{16.6} [10.5, 23.2] & \best{85.9}\\
\midrule
Qwen3-TTS & Base & \best{3.9} [2.0, 6.1] & \best{5.3} [2.5, 8.6] & 66.9 [62.8, 70.8] & 34.4\\
 & Short-SFT & 10.0 [7.1, 13.5] & 49.6 [37.1, 61.3] & 88.0 [85.5, 90.3] & 12.2\\
 & Long-SFT-Punct & 8.3 [5.3, 11.7] & 7.7 [5.6, 9.8] & \best{35.2} [27.0, 43.9] & \best{67.1}\\
\midrule
VoxCPM2 & Base & 12.4 [6.8, 18.7] & 14.2 [3.4, 29.4] & 94.9 [91.0, 98.0] & 5.2\\
 & Short-SFT & 4.4 [2.2, 6.9] & 4.1 [3.2, 4.9] & 90.5 [86.9, 93.9] & 10.5\\
 & Long-SFT-Punct & \best{3.4} [1.6, 5.7] & \best{2.4} [1.8, 3.0] & \best{49.0} [40.8, 56.7] & \best{54.9}\\
\midrule
F5-TTS & Base & 45.0 [34.7, 54.7] & 98.7 [97.7, 99.6] & 99.7 [99.7, 99.8] & 0.3\\
 & Short-SFT & 44.2 [34.9, 52.8] & 97.6 [96.9, 98.3] & 99.3 [99.2, 99.5] & 0.7\\
 & Long-SFT-Punct & \best{27.4} [20.9, 34.2] & \best{81.4} [79.5, 83.2] & \best{94.7} [94.4, 95.0] & \best{5.4}\\
\bottomrule
\end{tabular}
\end{table*}

\subsection{State-continuous inference and evaluation sets}

Continuous inference uses one target text and one voice-conditioning event,
followed by an unsplit autoregressive trajectory or full-sequence flow solve.
External splitting (including the CosyVoice3 text splitter), state resets,
best-of-$N$ selection and regeneration are disabled; all attempts are retained.
Context, output and acoustic-buffer limits are identical across the arms of a
backbone and documented with the code. They bind at about 1,200 words: VoxCPM2's
text fills about 4,900 of its 8,192 positions, so 16, 14 and 6 of 22 Base,
Short-SFT and Long-SFT-Punct attempts end at that limit. F5-TTS estimates
duration from prompt audio and text length; it has no end-of-sequence (EOS)
decision, and planned duration is not a learned stopping measure.

The two-voice set (Table~\ref{tab:bybucket}) combines six source texts
(four development recordings and two constructed stress texts), five nested
duration ranges and two held-out voices: 60 attempts per arm.
The published-text set (Table~\ref{tab:crossmodel}) uses 50 pairs from 22 works
and 21 reference speakers that contribute no fine-tuning data: eight LibriVox
readers from Russian LibriSpeech~\cite{openslr96ruls} and 13 corpus speakers, six
from test and seven from dev channels, first used in an earlier recorded-text pilot. Each work supplies one long passage near 1,200 normalized words, and 14 short
and 14 medium prefixes target 75 and 300 words (ranges 60--90, 240--360 and
960--1,440 words, not reference-audio durations). Inputs are literal published spans screened against training and prior
evaluation text; every arm uses the same pairs.

The 50 pairs are a deterministic, metadata-only subset of a planned 120-pair set
(Section~\ref{sec:limitations}). All input is punctuated and capitalized. Among the CosyVoice3 arms, only
Long-SFT-Punct matches this format in training; the two adapted arms of each
other backbone share one training text format.

\subsection{Evaluator and paired uncertainty}
\label{sec:evaluator}

GigaAM-v3-RNNT~\cite{kutsakov2025gigaam} transcribes every generated output.
Mean per-attempt WER counts substitutions, deletions and insertions relative to
reference words; correct-word recall is aligned correct words divided by
reference words. Unusable audio becomes an empty hypothesis.
The published-text set has no paired human recording, so it uses no human-ASR
floor or completion label.

Table~\ref{tab:bybucket} retains the earlier completion rule, frozen before SFT:
an autoregressive EOS before 95\% aligned text coverage is early; remaining
outputs are labeled degraded when their WER is at least 30 points above a
human-ASR floor or when they meet fixed repetition criteria. Watchdog stops retain
their label. Constructed texts lack
human recordings and skip the WER-floor test, so complete does not imply a
faithful transcript.

For the published-text set, each backbone's primary contrast compares its long and short
arms (Long-SFT for CosyVoice3, Long-SFT-Punct for the other backbones)
on the 22 longest inputs. Paired pointwise 95\% intervals use 10,000 resamples
of the 22 work clusters, retaining associated prefixes and voices; independently
resampling work and voice levels is a crossed-factor sensitivity check that widens
but does not overturn any primary interval (CosyVoice3: [$-67.7$, $-34.6$]).
Intervals condition on fitted checkpoints and inference seed and do not adjust
for multiple comparisons. Intervals for the two-voice set instead resample its
six source texts.

\section{Experiments}
\label{sec:experiments}

\subsection{Earlier controlled duration comparison}
\label{sec:duration}

Long-SFT improves continuous \model{} reading with two reference voices
(Table~\ref{tab:bybucket}).
Short-SFT resembles Base, whereas Long-SFT gains 45.0 completion points [33.3, 56.7] over Short-SFT under matched exposure.  Neither Base nor Short-SFT
completes an item beyond 90 s. The sentence-split control completes all items
at 3.84\% WER; our continuous protocol claims no superiority over it.  Repeated
decoding shifts bucket percentages by up to $\sim$20 points, so these single-seed
rows are descriptive.

Punctuated transcripts add a secondary gain: Long-SFT-Punct reaches 83.3\%
completion, $+11.7$ completion points and $-6.11$ WER points over Long-SFT
(Table~\ref{tab:bybucket}), but this comparison changes both transcript source
and formatting.  Giving Long-SFT lowercased, unpunctuated input
matched to its training format yields 88.3\% completion and 10.07\% WER, which
Long-SFT-Punct (83.3\%, 14.36\%) does not improve on.  The gain is therefore
consistent with format matching; a separate benefit of punctuation content remains
unestablished on this single-seed two-voice set.  Three preregistered mixtures with
external short clips did not improve Long-SFT and increased repetition.

\subsection{Content fidelity across backbones}
\label{sec:crossmodel}

The completed published-text comparison tests the training views on new source
passages across four backbones (Table~\ref{tab:crossmodel}). For the three
autoregressive models, Long-minus-Short WER changes at approximately 1,200 words,
in percentage points (pp), are CosyVoice3: $-52.7$ pp [$-62.7$, $-42.2$]; Qwen3-TTS:
$-52.7$ pp [$-61.7$, $-43.4$]; VoxCPM2: $-41.5$ pp [$-47.7$, $-35.4$].
Negative changes indicate fewer word errors. All are paired effects on the same
22 work passages, not 50 independent attempts. The gap opens well before that
length: at about 300 words CosyVoice3 Base and Short-SFT already reach 84.8\% and
82.2\% WER, while Long-SFT-Punct stays at 5.6\%; at 1,200 words it improves on Base
by 83.2 pp [$-89.3$, $-76.6$]. Against Base rather than Short-SFT
the changes are $-52.5$, $-31.6$ and $-46.0$ pp, and at about 75 words Qwen3-TTS
Long-SFT-Punct is 4.4 pp worse than Base [1.9, 7.2], so the benefit is specific
to long inputs. The released stop logs show how arms fail at 1,200 words. Every
CosyVoice3 Base and Short-SFT attempt emits EOS early (median 9.7 s and 5.3 s of
audio), and Qwen3-TTS Short-SFT stops after a median 60 s, consistent with windows
teaching a stopping decision near the window length; this is how Short-SFT can be
worse than Base (Qwen3-TTS: 88.0\% vs 66.9\% WER).

\subsection{Flow matching and inference protocol}
\label{sec:f5}

F5-TTS extends the paired comparison to a flow-matching model. Its longest-input
Long-minus-Short WER change is $-4.6$ pp [$-4.9$, $-4.4$], but its absolute WER in
Table~\ref{tab:crossmodel} shows that a relative change alone does not establish
reliable long-text reading. The duration schedule is shared across checkpoints,
so this comparison evaluates
content realization rather than an EOS decision. Its largest gain is on short
inputs: at about 75 words Long-SFT-Punct lowers WER from 44.2\% to 27.4\%.

Chunking provides a distinct practical reference. On the \emph{earlier} recorded-text
53-pair set, F5-TTS Base WER was 3.58\% with sentence chunking and
73.35\% with continuous synthesis. Those values use different texts from
Table~\ref{tab:crossmodel}, are not rows of the published-text comparison, and
establish no published-text ranking.

\subsection{Evaluation checks and practical limits}

The published-text WER uses one RNNT recognizer; checks with a second recognizer
earlier supported the CosyVoice3 ranking on recorded texts.
Figure~\ref{fig:qualitytime} adds windowed DistillMOS~\cite{stahl2025distillation}
and WeSpeaker~\cite{wang2022wespeaker} measurements on the published-text outputs.
F5-TTS Long-SFT-Punct has higher DistillMOS than Base despite high WER; speaker similarity declines over time for Qwen3-TTS and VoxCPM2 but stays flat
for CosyVoice3 (Fig.~\ref{fig:qualitytime}b). Curves condition on scorable windows, with support shown below. Truncated outputs can score well, so these
proxies complement all-attempt WER and do not replace listening tests.

\section{Limitations}
\label{sec:limitations}

The published-text set covers 22 literary works and 21 voices, and its 50 pairs
were chosen from a planned 120 by a metadata-only rule after preliminary results
had been seen. GigaAM supplies training transcripts and scores outputs, and
no listening study was run. Transcript-source, formatting and VoxCPM2
corpus-composition effects are not isolated, and one F5-TTS model does not
represent all non-autoregressive systems. Corpus transcripts and speaker labels
are automatic.

\section{Conclusion}
\label{sec:conclusion}

\corpus{} preserves long continuous Russian speech with paired short-window
training views. Across CosyVoice3, Qwen3-TTS, VoxCPM2 and F5-TTS, a two-voice
study and a published-text comparison report absolute content fidelity and paired
Short/Long changes; the intervals favor long targets for the three autoregressive
backbones and by a small margin for F5-TTS, whose reading remains unreliable. The corpus supports studying sustained reading
while keeping content fidelity and
perceptual quality distinct.

\medskip\noindent\textbf{Acknowledgments.} Claude Fable 5.1 (Anthropic) was
used to polish the wording; all technical content, experiments, analyses and
conclusions are the authors' own, who reviewed and edited all text.
\label{sec:contentend}

\clearpage
\section{Compliance with Ethical Standards}

This research study was conducted retrospectively using speech recordings that
their uploaders published on YouTube with a Creative Commons Attribution
declaration.  No recordings were made for this work and no experiment involved
human participants, so ethical approval was not required.  The uploader's
declaration is license evidence; it is not an independent audit of rights in the
underlying texts or performances and does not establish speakers' consent to use
in speech synthesis.  The corpus is intended for research on long-form synthesis.  \corpus{} is distributed
through \datasetlink{} with per-source provenance and license evidence.
Original recordings remain governed by their respective source licenses; the
release does not relicense third-party material.  Attribution corrections and
removal requests are handled through the repository's discussion page and the
contact address given above.  The authors declare no conflicts of interest.

\bibliographystyle{IEEEbib}
\bibliography{refs}

@article{du2025cosyvoice3,
  title   = {{CosyVoice 3}: Towards In-the-Wild Speech Generation via Scaling-Up and Post-Training},
  author  = {Du, Zhihao and Gao, Changfeng and Wang, Yuxuan and Yu, Fan and Zhao, Tianyu and others},
  journal = {arXiv preprint arXiv:2505.17589},
  year    = {2025}
}

@misc{qwen2025cosyvoicecode,
  title        = {{CosyVoice}: Multi-Lingual Large Voice Generation Model},
  author       = {{Qwen Audio Team}},
  year         = {2025},
  howpublished = {\url{https://github.com/QwenAudio/CosyVoice}},
  note         = {Accessed 30 Aug. 2026}
}

@article{du2024cosyvoice2,
  title   = {{CosyVoice 2}: Scalable Streaming Speech Synthesis with Large Language Models},
  author  = {Du, Zhihao and Wang, Yuxuan and Chen, Qian and Shi, Xian and Lv, Xiang and others},
  journal = {arXiv preprint arXiv:2412.10117},
  year    = {2024}
}

@article{hu2026qwen3tts,
  title   = {{Qwen3-TTS} Technical Report},
  author  = {Hu, Hangrui and Zhu, Xinfa and He, Ting and Guo, Dake and Zhang, Bin and others},
  journal = {arXiv preprint arXiv:2601.15621},
  year    = {2026}
}

@misc{qwen2026qwen3ttscode,
  title        = {{Qwen3-TTS}: Official Models and Fine-Tuning Code},
  author       = {{Qwen Team}},
  year         = {2026},
  howpublished = {\url{https://github.com/QwenLM/Qwen3-TTS}},
  note         = {Accessed 31 Aug. 2026}
}

@article{zhou2026voxcpm2,
  title   = {{VoxCPM2} Technical Report},
  author  = {Zhou, Yixuan and Zeng, Guoyang and Liu, Xin and Li, Xiang and Yu, Renjie and others},
  journal = {arXiv preprint arXiv:2606.06928},
  year    = {2026}
}

@misc{openbmb2026voxcpmcode,
  title        = {{VoxCPM}: Tokenizer-Free {TTS} for Multilingual Speech Generation, Creative Voice Design, and True-to-Life Cloning},
  author       = {{OpenBMB}},
  year         = {2026},
  howpublished = {\url{https://github.com/OpenBMB/VoxCPM}},
  note         = {Accessed 2 Sept. 2026}
}

@article{ghosh2026magpiettslf,
  title   = {{MagpieTTS-LF}: Inference-Time Long-Form Speech Generation Without Training on Long-Form Data},
  author  = {Ghosh, Subhankar and Li, Jason and Neekhara, Paarth and Hussain, Shehzeen and Langman, Ryan and others},
  journal = {arXiv preprint arXiv:2606.18485},
  year    = {2026}
}

@inproceedings{peng2026voicestar,
  title     = {{VoiceStar}: Robust Zero-Shot Autoregressive {TTS} with Duration Control and Extrapolation},
  author    = {Peng, Puyuan and Zheng, Zhisheng and Li, Shang-Wen and Mohamed, Abdelrahman and Harwath, David},
  booktitle = {Findings of the Association for Computational Linguistics: ACL 2026},
  pages     = {11737--11754},
  year      = {2026},
  doi       = {10.18653/v1/2026.findings-acl.570}
}

@inproceedings{pan2026lfsbench,
  title     = {Comprehensive Benchmarking of Long-Form Speech Generation in Diverse Scenarios},
  author    = {Pan, Changhao and Yang, Rui and Wang, Han and Zhou, Zhuan and He, Xuming and others},
  booktitle = {Findings of the Association for Computational Linguistics: ACL 2026},
  pages     = {2365--2400},
  year      = {2026},
  doi       = {10.18653/v1/2026.findings-acl.112}
}

@article{clark2019evaluating,
  title   = {Evaluating Long-Form Text-to-Speech: Comparing the Ratings of Sentences and Paragraphs},
  author  = {Clark, Rob and Silen, Hanna and Kenter, Tom and Leith, Ralph},
  journal = {arXiv preprint arXiv:1909.03965},
  year    = {2019}
}

@article{kutsakov2025gigaam,
  title   = {{GigaAM}: Efficient Self-Supervised Learner for Speech Recognition},
  author  = {Kutsakov, Aleksandr and Maximenko, Alexandr and Gospodinov, Georgii and Bogomolov, Pavel and Minkin, Fyodor},
  journal = {arXiv preprint arXiv:2506.01192},
  year    = {2025}
}

@inproceedings{fiscus1997rover,
  title     = {A Post-Processing System to Yield Reduced Word Error Rates: Recognizer Output Voting Error Reduction ({ROVER})},
  author    = {Fiscus, Jonathan G.},
  booktitle = {Proc. IEEE ASRU},
  year      = {1997},
  doi       = {10.1109/ASRU.1997.659110}
}

@inproceedings{wang2022wespeaker,
  title     = {{WeSpeaker}: A Research and Production Oriented Speaker Embedding Learning Toolkit},
  author    = {Wang, Hongji and Liang, Chengdong and Wang, Shuai and Chen, Zhengyang and Zhang, Binbin and others},
  booktitle = {Proc. ICASSP},
  year      = {2023}
}

@article{borodin2026balalaika,
  title   = {Balalaika: Data-Centric, Prosody-Aware Annotation Pipeline for {Russian} Speech},
  author  = {Borodin, Kirill and Vasiliev, Nikita and Kudryavtsev, Vasiliy and Maslov, Maxim and Gorodnichev, Mikhail and Mkrtchian, Grach},
  journal = {arXiv preprint arXiv:2507.13563},
  year    = {2025},
  url     = {https://arxiv.org/abs/2507.13563}
}

@article{koizumi2023librittsr,
  title   = {{LibriTTS-R}: A Restored Multi-Speaker Text-to-Speech Corpus},
  author  = {Koizumi, Yuma and Zen, Heiga and Karita, Shigeki and Ding, Yifan and Yatabe, Kohei and others},
  journal = {arXiv preprint arXiv:2305.18802},
  year    = {2023},
  url     = {https://arxiv.org/abs/2305.18802}
}

@article{langman2025hifitts2,
  title   = {{HiFiTTS-2}: A Large-Scale High Bandwidth Speech Dataset},
  author  = {Langman, Ryan and Yang, Xuesong and Neekhara, Paarth and Hussain, Shehzeen and Casanova, Edresson and others},
  journal = {arXiv preprint arXiv:2506.04152},
  year    = {2025},
  url     = {https://arxiv.org/abs/2506.04152}
}

@article{gabdrakhmanov2019ruslan,
  title   = {{RUSLAN}: Russian Spoken Language Corpus for Speech Synthesis},
  author  = {Gabdrakhmanov, Lenar and Garaev, Rustem and Razinkov, Evgenii},
  journal = {arXiv preprint arXiv:1906.11645},
  year    = {2019},
  url     = {https://arxiv.org/abs/1906.11645}
}

@article{chen2024f5tts,
  title   = {{F5-TTS}: A Fairytaler that Fakes Fluent and Faithful Speech with Flow Matching},
  author  = {Chen, Yushen and Niu, Zhikang and Ma, Ziyang and Deng, Keqi and Wang, Chunhui and others},
  journal = {arXiv preprint arXiv:2410.06885},
  year    = {2024},
  url     = {https://arxiv.org/abs/2410.06885}
}

@misc{chen2026f5code,
  title        = {{F5-TTS}: Official Inference Code},
  author       = {{F5-TTS Contributors}},
  year         = {2026},
  howpublished = {\url{https://github.com/SWivid/F5-TTS}},
  note         = {Accessed 10 Sept. 2026}
}

@inproceedings{stahl2025distillation,
  title     = {Distillation and Pruning for Scalable Self-Supervised Representation-Based Speech Quality Assessment},
  author    = {Stahl, Benjamin and Gamper, Hannes},
  booktitle = {Proc. ICASSP},
  year      = {2025},
  doi       = {10.1109/ICASSP49660.2025.10888007}
}

@inproceedings{park2025speechssm,
  title     = {Long-Form Speech Generation with Spoken Language Models},
  author    = {Park, Se Jin and Salazar, Julian and Jansen, Aren and Kinoshita, Keisuke and Ro, Yong Man and Skerry-Ryan, RJ},
  booktitle = {Proc. ICML},
  pages     = {48245--48261},
  year      = {2025}
}

@inproceedings{kang2024libriheavy,
  title     = {Libriheavy: A 50,000 Hours {ASR} Corpus with Punctuation Casing and Context},
  author    = {Kang, Wei and Yang, Xiaoyu and Yao, Zengwei and Kuang, Fangjun and Yang, Yifan and others},
  booktitle = {Proc. ICASSP},
  year      = {2024}
}

@misc{openslr96ruls,
  title        = {{Russian LibriSpeech (RuLS)}},
  howpublished = {OpenSLR resource SLR96, \url{https://www.openslr.org/96/}},
  note         = {Accessed 15 Sept. 2026}
}

@misc{misha2025f5ru,
  title        = {{F5-TTS\_RUSSIAN}: Community {Russian} {F5-TTS} Checkpoints},
  author       = {{Misha24-10}},
  year         = {2025},
  howpublished = {\url{https://huggingface.co/Misha24-10/F5-TTS_RUSSIAN}},
  note         = {Accessed 15 Sept. 2026}
}

@inproceedings{zhao2024mossformer2,
  title     = {{MossFormer2}: Combining Transformer and {RNN}-Free Recurrent Network for Enhanced Time-Domain Monaural Speech Separation},
  author    = {Zhao, Shengkui and Ma, Yukun and Ni, Chongjia and Zhang, Chong and Wang, Hao and others},
  booktitle = {Proc. ICASSP},
  pages     = {10356--10360},
  year      = {2024},
  doi       = {10.1109/ICASSP48485.2024.10445985}
}

@inproceedings{li2023yodas,
  title     = {{YODAS}: {YouTube}-Oriented Dataset for Audio and Speech},
  author    = {Li, Xinjian and Takamichi, Shinnosuke and Saeki, Takaaki and Chen, William and Shiota, Sayaka and Watanabe, Shinji},
  booktitle = {Proc. IEEE ASRU},
  year      = {2023}
}

\end{document}